%% file: manuscript.tex
\documentclass[sigconf]{acmart}

\AtBeginDocument{%
  \providecommand\BibTeX{{%
    \normalfont B\kern-0.5em{\scshape i\kern-0.25em b}\kern-0.8em\TeX}}}

\copyrightyear{2023}
\acmYear{2023}
\setcopyright{rightsretained}
\acmConference[DIS '23]{Designing Interactive Systems Conference}{July 10--14, 2023}{Pittsburgh, PA, USA}
\acmBooktitle{Designing Interactive Systems Conference (DIS '23), July 10--14, 2023, Pittsburgh, PA, USA}
\acmDOI{10.1145/3563657.3596101}
\acmISBN{978-1-4503-9893-0/23/07}

\usepackage{graphicx}  
\usepackage{url}
\usepackage{booktabs}  
\usepackage{caption}
\usepackage{subcaption}
\usepackage{tabularx}
\usepackage{tablefootnote}

\graphicspath{{figures/}{pictures/}{images/}{./}}

\begin{document}

\title[How Do UX Practitioners Communicate AI as a Design Material?]{How Do UX Practitioners Communicate AI as a Design Material? Artifacts, Conceptions, and Propositions}


\author{K. J. Kevin Feng}
\affiliation{%
  \institution{University of Washington}
  \city{Seattle}
  \country{USA}}
\email{kjfeng@uw.edu}

\author{Maxwell James Coppock}
\affiliation{%
  \institution{University of Washington}
  \city{Seattle}
  \country{USA}}
\email{mxcopp97@uw.edu}

\author{David W. McDonald}
\affiliation{%
  \institution{University of Washington}
  \city{Seattle}
  \country{USA}}
\email{dwmc@uw.edu}

\renewcommand{\shortauthors}{Feng et al.}

\begin{abstract}
UX practitioners (UXPs) face novel challenges when working with and communicating artificial intelligence (AI) as a design material. We explore how UXPs communicate AI concepts when given hands-on experience training and experimenting with AI models. To do so, we conducted a task-based design study with 27 UXPs in which they prototyped and created a design presentation for a AI-enabled interface while having access to a simple AI model training tool. Through analyzing UXPs' design presentations and post-activity interviews, we found that although UXPs struggled to clearly communicate some AI concepts, tinkering with AI broadened common ground when communicating with technical stakeholders. UXPs also identified key risks and benefits of AI in their designs, and proposed concrete next steps for both UX and AI work. We conclude with a sensitizing concept and recommendations for design and AI tools to enhance multi-stakeholder communication and collaboration when crafting human-centered AI experiences. 
\end{abstract}

\ccsdesc[500]{Human-centered computing~Empirical studies in HCI}
\ccsdesc[500]{Human-centered computing~Interaction design process and methods}

\keywords{design communication, user experience, artificial intelligence}

\maketitle

\section{Introduction}
\input{./sections/intro}

\section{Related Work}
\input{./sections/related-work}

\section{Methods}
\input{./sections/methods}

\section{Results}
\input{./sections/results}

\section{Discussion}
\input{./sections/discussion}

\section{Limitations and Future Work}
\input{./sections/limitations-fw}

\section{Conclusion}
\input{./sections/conclusion}


\begin{acks}
We extend a warm thanks to all of our participants and anonymous reviewers. We would also like to thank Amodini Khade and Katrina Vergara for insightful discussions and assistance with data analysis.
\end{acks}

\bibliographystyle{ACM-Reference-Format}
\bibliography{refs}





\end{document}

%% file: sections/intro.tex
For user experience practitioners\footnote{We use ``user experience practitioners'' in this paper as a general term referring to both UX designers and researchers (and managers thereof).} (UXPs), communicating design work is a critical component of the design process. UXPs often work iteratively and create multiple variations of a design before meeting with stakeholders---including clients, product managers, and fellow UXPs---to discuss and improve their work \cite{oleary2018charrette}. These review sessions, known as \textit{design critiques}, are central to conveying design knowledge \cite{blevis2007studio}. They often involve practitioners bringing low-fidelity sketches into a common space for general communication of concepts and rapid elimination of bad ideas, taking turns to review and provide constructive criticism to each others' work, and collectively brainstorming improvements for future iterations \cite{reimer2003teaching}. In addition to design critiques, a more recent collaborative practice that arose with the ubiquity of interactive digital products is the \textit{developer handoff} between UXPs and software developers. Handoffs can be a common source of collaborative breakdowns between UXPs and software developers as fundamental differences exist between visual representations of interfaces (which UXPs work with) and code-based ones (which developers work with) \cite{maudet2017, leiva2019, feng2023collaboration}. To more effectively design for digital technologies, the design community has argued for framing technology as a \textit{design material} to capture vital designerly properties that are shared across physical and computational materials \cite{robles2010material}.

In the modern technological landscape, advancements in artificial intelligence (AI) propels its incorporation into user-facing technology. This presents novel opportunities for UXPs to engage with AI as a design material \cite{holmquist2017intelligence}. However, prior work has shown that UXPs encounter numerous novel challenges when designing with AI that emerge from issues including understanding AI models' capabilities and limitations \cite{yang2020difficult, dove2017material}, calibrating user trust \cite{beaudouin2009prototyping, benjamin2021material}, mitigating potentially harmful model outputs \cite{holmquist2017intelligence, lee2020fair}, a lack of model explainability \cite{zhou2020ml, cai2021onboarding, yang2018imt}, and unfamiliarity with data science concepts \cite{yang2018experienced}. A more recent thread of work has investigated tools and process models to mitigate some of the well-documented challenges \cite{subramonyam2021protoai, subramonyam2021towards}, as well as collaborative artifacts in enterprise settings to support UXPs in envisioning possibilities for AI-enabled designs \cite{yildirim2022enterprise}. New tools for AI non-experts have also made experimentation with AI, and in particular machine learning, as a design material more approachable \cite{carney2020teachable, lobe, liner}. 

Given the increased attention paid by researchers and practitioners in this area, surprisingly little work has situated UXPs' experiences communicating AI as a design material in critique-like settings. We posit that inability to do so hinders our understanding of how to better support UXPs when working with AI in two main ways. First, critiques are deeply embedded in design practice and are key to fostering shared understanding and collaboration \cite{reimer2003teaching}, especially in human-centered AI workflows where success is contingent on multiple groups of experts working together \cite{subramonyam2022leaky}. Second, while prior work has surfaced boundary objects used by UXPs and technical AI collaborators, the goal of those objects may be different ones created for a critique. Some artifacts---such as probes to understand model performance on specific user data--- are fit for resolving specific implementation concerns and edge cases, while others---such as interface sketches---serve to solicit feedback on design feasibility and next steps for iteration. Additionally, these artifacts may vary in nature depending on the availability of AI experts to explain to UXPs how the model works, which prior work has shown to be scarce \cite{yang2018experienced}. We eliminate this variable by providing UXPs with a tool to explore and experiment with the properties of AI throughout their design workflows.

In this work, we conducted a contextual inquiry in the form of a task-based design study with 27 UXPs in industry to address these gaps. We prompted them to create low-fidelity representations of a AI-enabled image classification along with a design presentation for stakeholders to communicate their ideas. They did so while having access to Google's Teachable Machine \cite{carney2020teachable}, a simple web-based AI model training tool, to tinker with AI as a design material. We conducted post-activity interviews and analyzed their design presentations, finding that interactive exploration of AI expanded common ground between UXPs and their collaborators for communicating AI-enabled designs. UXPs were also able to raise concrete benefits and drawbacks of incorporating AI into their work, as well next steps for iteration, in their design presentations. However, they still struggled to effectively communicate, partly due to knowledge gaps, and partly due to fundamental differences in evaluating AI's success. In light of our findings, we borrow the notion of \textit{fidelity} from prototyping and apply it to AI models as a sensitizing concept for UXPs when designing with AI. We also see potential in transforming AutoML tools into \textit{collaborative platforms} for interdisciplinary AI teams to share artifacts and knowledge, and propose \textit{probabilistic user flows} as a way to bridge user-centered and technical evaluations of AI.

In sum, this work makes the following key contributions:
\begin{itemize}
    \item Insights (both conceptual and graphical) about how UXPs communicate AI as a design material to their stakeholders, through an analysis of design presentations and accompanying interviews from 27 UXPs in industry.
    \item A sensitizing concept for UXPs when engaging with AI as a design material.
    \item Design recommendations for AI and UX tools to better support interdisciplinary teams in developing AI user experiences.
\end{itemize}

%% file: sections/related-work.tex
\subsection{UX Challenges of Working with AI as a Design Material}
\label{s:rr-ml-material}
In his 2003 article \textit{On Materials}, Doordan \cite{doordan2003materials} introduces the material framework for designers to highlight the critical inclusion of materiality in discussions of designed artifacts. The framework consists of 3 main parts. \textit{Fabrication} is the processing of materials (extraction, refinement, and preparation) for initial use. \textit{Application} is the transformation by which the processed materials are turned into products. \textit{Appreciation} is the reception of and discourse around the material by communities of individuals who interact with it (typically via a product). 

Doordan's examples from his framework were centered on tangible products, but prior work in the design and HCI community has argued for the importance of applying the framework to understand the design and use of rapidly advancing digital technology \cite{hondros2015internet, robles2010material}. Specifically, researchers have framed AI as a design material to expose unique difficulties for designers working with AI \cite{dove2017material, yang2018material, luciani2018material, benjamin2021material, holmquist2017intelligence, yildirim2022enterprise}. Through the lens of Doordan's material framework, we observe that the detachment of designers from the \textit{fabrication} of the material can trigger a cascade of subsequent challenges. The material---in the form of AI models, infrastructure, and data---are often prepared by AI teams (data scientists, AI engineers, etc.) who are far removed from user-centered design principles that ground designers' work \cite{yildirim2022enterprise, subramonyam2022leaky, yang2018material}. As a result, designers work with AI without sufficient understanding of its capabilities and limitations---e.g., a blackbox \cite{zhou2020ml}. Designers are then easily surprised by AI's stochastic behaviours and errors, limiting their abilities to properly calibrate the material when generating design solutions \cite{beaudouin2009prototyping, yang2020difficult, benjamin2021material} and account for unintended ethical issues \cite{holmquist2017intelligence, lee2020fair}. Lastly, it is challenging for designers to collect \textit{appreciation}---that is, holistically evaluate AI-enabled designs with users. The material itself may constantly evolve new data and user inputs, and new AI capabilities may be unlocked sporadically \cite{yang2018material, yang2020difficult}. 


To tackle these challenges, researchers and practitioners have established human-AI guidelines to offer both cognitive scaffolding and educational materials when designers work with AI \cite{pair-guidebook, hax-guidelines, apple-guidelines}. Researchers have also proposed tools that combine UI prototyping with AI model exploration \cite{subramonyam2021protoai}, process models \cite{zhou2020ml, subramonyam2021towards} and boundary representations \cite{yildirim2022enterprise, subramonyam2022leaky} for human-centered AI teams, and metaphors for generative probing of models \cite{dove2020monsters}. In particular, Subramonyam et al. \cite{subramonyam2022leaky} found that UX and AI practitioners use low-level representations which they call \textit{leaky abstractions} (e.g., low-fidelity prototypes and raw data) to communicate across domain boundaries. They also introduce \textit{data probes} \cite{subramonyam2021towards} as design probes enabled by user data to identify desirable AI characteristics and user experiences. 

One key assumption behind solutions derived from prior work is that designers do not have the agency to manipulate AI directly as a design material---their understanding of AI comes from relying on explanations from AI experts. Drawing once again from Doordan's material framework, this still does not resolve the foundational issue in the \textit{fabrication} step: those who applying the material to a product are not the ones who prepare the material. In our study, we remove this assumption and allow designers to craft their own AI models to personally experience AI's materiality via tinkering. We then examine the low-level representations they create as a result of those experiences to discuss opportunities for enhanced communication strategies and tools in the human-centered AI design process.

\subsection{Collaboration Between UXPs and Technical Stakeholders}
Hollan, Hutchins, and Kirsh developed the theory of distributed cognition out of the observation that information is no longer confined to the individual, but traverses fluidly across individuals, teams, and communities \cite{hollan2000}. This is especially true when designing user-facing AI systems. The AI experts who are wrangling data and building models are not trained in identifying user needs, while UXPs who center their work in user-centered design are unfamiliar with AI concepts \cite{subramonyam2022leaky}. Both groups of individuals are often not the domain experts (e.g., doctors) who deeply understand contextual nuances in situations where the AI system will reside \cite{ramos2020imt}. As such, there is increasing interest in techniques and artifacts to bridge distributed expertise in human-centered AI \cite{hong2021playbook, holstein2019fairness, passi2018trust, cai2019medical, zhang2020data}. 

Boundary objects \cite{star1989} have been shown by prior work to provide critical support in establishing shared understanding and collaborative workflows in interdisciplinary AI teams \cite{yang2019sketching, john2004boundary, cai2021onboarding}. Examples used by UXPs to work with AI experts include data visualizations, wireframes, and design annotations \cite{yang2018experienced}. Other artifacts for collaboration include boundary negotiating artifacts \cite{lee2007} and coordinative artifacts \cite{schmidt2002}. Boundary negotiating artifacts differ from boundary objects in that they 1) operate outside the realm of standard, well-defined processes, and 2) aim to negotiate and shift domain boundaries instead of aiming to increase the permeability of static ones \cite{lee2007}. Coordinative artifacts are inherently effective at coordinating activities between diverse groups, but those groups may not necessarily need to work across or negotiate boundaries \cite{schmidt2002}. In this work, we primarily use ``boundary objects'' to describe artifacts for collaboration as boundary negotiation is not a primary goal. Furthermore, although the artifacts are often effective  facilitators of coordination, their primary value lies in their ability to communicate information across boundaries. 


In our study, we provide a glimpse into what the human-centered AI design process may look like if it followed the traditional developer handoff workflow: UXPs take the lead by ``handing off'' low-fidelity AI-enabled designs to technical collaborators for discussion and refinement \cite{handoff-overview}. We do so in an attempt to capture the ability of handoffs to better align products with user needs: user research and preliminary feedback gathering could occur \textit{before} the product was formally built \cite{handoff-overview}. However, unlike a traditional handoff, we do expect these early design explorations to be implemented directly. Rather, we view these designs as potential boundary objects with the ability to foster interdisciplinary collaboration and embody distributed cognition across practitioner communities.

\subsection{Low- and No-Code AI Tools}
As the field of AI advances, so does interest in making the technology more accessible beyond a small group of technical experts. Interactive machine learning (IML) is a paradigm that aims to provide users or user groups with an active role in the model building process with rapid, iterative cycles of information input and review \cite{dudley2018iml, amershi2014power}. Fails and Olsen describes this involvement of human guidance in the AI workflow as ``human-in-the-loop,'' \cite{fails2003iml}, a term that has since become ubiquitous in AI and HCI \cite{mosqueira2022human}. For instance, in active learning, a model dynamically queries a human for data labels and leverages that feedback to improve performance while reducing the need for pre-labelled examples \cite{settles2012active, chao2010transparent, monarch2021human, raghavan2006active, settles2011theories}. Researchers have also integrated human-in-the-loop concepts into transfer learning systems---users can interactively transfer learned representations from a large, generalized model to a smaller, domain-specific model \cite{mishra2021transfer, kumaraswamy2020interactive}. While active learning and transfer learning primarily rely on an automated agent to extract representations from data, interactive machine teaching \cite{ramos2020imt} seeks to allow the user to provide learnable representations instead. Through an iterative process of \textit{planning} a teaching task and gathering the required materials (e.g., data), \textit{explaining} concepts to the machine learner through interface affordances such as text highlighting, and \textit{reviewing} the learner's progress and making any corrections or updates as necessary \cite{ng2020imt, sanchez2022teaching, weitekamp2020imt, zhou2022gesture, yang2018imt}. Regardless of the approach or technique used by the system, human supervision---whether it be through data labelling, error correction, or concept teaching---stands as a vital backbone of IML.

Within IML, there has been developing interest in the usability of automated machine learning (AutoML) systems \cite{karmaker2021automl, xin2021whither, crisan2021automl}. AutoML systems are designed to automate various parts of the machine learning workflow (e.g., feature engineering \cite{liu2020autofis, severyn2013automatic}, model selection \cite{laredo2020automatic, truong2019towards}, hyperparameter tuning \cite{fusi2018probabilistic}) in an effort to reduce tedious work and increase efficiency \cite{nagarajah2019review}. Industry cloud-baesd offerings of AutoML systems such as Microsoft Azure's AutoML \cite{ms-aml} and IBM Watson's AutoAI \cite{ibm-aml} (among others, e.g., \cite{google-aml, ibm-aml, databricks-aml, amazon-aml}) claim that these systems can enable technical AI non- and semi-experts to create deployable AI models with little to no code. However, interviews with practitioners who used AutoML in their work revealed that most users are still expert data scientists, and those experts are concerned about non-experts' use of AutoML due to the potential to automate and amplify errors \cite{xin2021whither, crisan2021automl}. Besides major cloud-based AutoML offerings---which are meant to build robust, fully fledged models \cite{google-aml}---a few tools offer beginner friendly, no-code interfaces for non-experts to tinker with AI \cite{lobe, liner, carney2020teachable}. These tools, such as Google's Teachable Machine \cite{carney2020teachable}, are designed more for education than ensuring model robustness. 

In our study, we specifically examine how Teachable Machine can allow UXPs to experiment with and communicate AI as a design material. We notice that while current IML and AutoML tools are discussed in data science and educational contexts, surprisingly little work has provided concrete empirical evidence of how they may aid UXPs amidst the challenges of AI design mentioned in Section \ref{s:rr-ml-material}. Our work seeks to close this gap. 

%% file: sections/methods.tex
Our goal was to explore how UXPs communicated AI as a design material to other stakeholders on their team in a critique-like setting when developing a new user-facing, AI-enabled technology. To this end, we conducted a contextual inquiry \cite{beyer1999contextual} of UX practice to simulate the type of work UXPs may do in their natural work environment. We conducted this study virtually over Zoom to broaden our participant pool. While Beyer and Holtzblatt emphasized that contextual inquiry should be performed in users' natural environments \cite{beyer1999contextual} (in the case of UXPs, this would be their office), we note that many UXPs may work from home offices due to the popularity of remote and hybrid work, so we do not consider our virtual study to be against the spirit of contextual inquiry.

Since our primary focus was to investigate the communication of AI as a design material, we did not design our study to address on the related yet orthogonal question of whether AI is the \textit{ideal design material} to use for a particular scenario. We invite interested readers to refer to literature on ideation \cite{liao2023designerly} and low-fidelity prototyping \cite{feng2023iml} for insights on the latter.

\subsection{Participants}
We recruited 27 participants through various channels such as UX- and HCI-related Slack workspaces, mailing lists associated with our institution, UX professional groups on Discord, LinkedIn, and Twitter, and forwarding the study invite to personal connections. Eligible participants were those who had at least one year of professional work experience in UX and were currently employed as a UX professional at the time of the study. Participants were able to sign up on a rolling basis while we conducted the study until we reached saturation. We provided each participant with an honorarium of a \$40 Amazon gift card for completing the study. Our study was approved by our institution's IRB under ID STUDY00015065.

Out of our participants, 25 of them were based in the United States, one was in Europe, and one was in Asia. 21 identified as UX designers and 6 as UX researchers. In terms of professional full-time work experience in UX, 14 had 1-2 years, 8 had 3-5 years, 2 had 6-10 years, and 3 had over 11 years. The organizations our participants worked for were diverse, with 15 working for large companies with 1000+ employees, 5 in medium organizations with 201-1000 employees and 7 in small companies with less than 200 employees. The participants also had varied educational backgrounds, with 14 coming from visual/industrial design, 8 from computing, 7 from social \& behavioural sciences, 4 from management, 3 from natural sciences \& math, 2 from architecture, 2 from humanities, and 1 from informatics.

Only 7 of our 27 participants reported prior experience designing with AI. However, slightly over half (15) reported they had previous exposure to AI through various avenues such as employer workshops, university courses, and online tutorials. 

\subsection{Study Structure}

We conducted 1:1 study sessions with our 27 participants. Each session was 2 hours in length and consisted of three main parts: Tutorial, Activity, and Interview. The first 90 minutes was dedicated to the Tutorial and Activity, while the Interview occupied the remaining 30 minutes.

\subsubsection{Tutorial}
To start the study, participants downloaded a folder prepared by the research team containing all necessary materials. Participants were then guided by a member of the research team through a tutorial of Teachable Machine, a web-based interactive tool for training AI models. Participants trained two simple image classification models to classify breeds of dogs to get acquainted with the tool: a binary classification model and a 3-class model. We provided all training images, which we sourced from Wikimedia Commons \cite{wikicommons}, in the folder they downloaded. Participants then evaluated the models they trained by uploading additional images of dogs (which we also provided) and checking if the model was correct in its classification. All participants stated they were comfortable using Teachable Machine to train image classification models by the end of the tutorial. The tutorial typically lasted 10 minutes.

We selected Teachable Machine as our AI tool of choice for this study for its easy setup, simple interface, and tight feedback loop. Because Teachable Machine is free and publicly accessible, it allowed participants to start tinkering without the need to download any software or create user accounts. Additionally, since most UXPs are not well-versed in AI terminology, we appreciated that Teachable Machine's simple layout and drag-and-drop features significantly lowered the barrier for AI non-experts to start training AI\footnote{More specifically, Teachable Machine trains machine learning (ML) models. We use AI as a general term that encompasses ML and other implementation methods such as decision trees, while keeping in mind that ML is the most prevalent method used in modern AI systems \cite{russell2010artificial}.} models. Lastly, and perhaps most importantly, Teachable Machine offered the ability to quickly train, test, and iterate AI models. This was essential for rapid prototyping and has been acknowledged as a key consideration for AI interface design tools by prior work \cite{yang2020difficult, subramonyam2021protoai}. We considered and experimented with other no-code AI platforms, including AutoML offerings from Google, Microsoft, and IBM, but found that only Teachable Machine offered a tight feedback loop that is necessary for participants to be able to complete the study within a reasonable timeframe.

\begin{figure*}[h]
    \centering
    \includegraphics[width=1\textwidth]{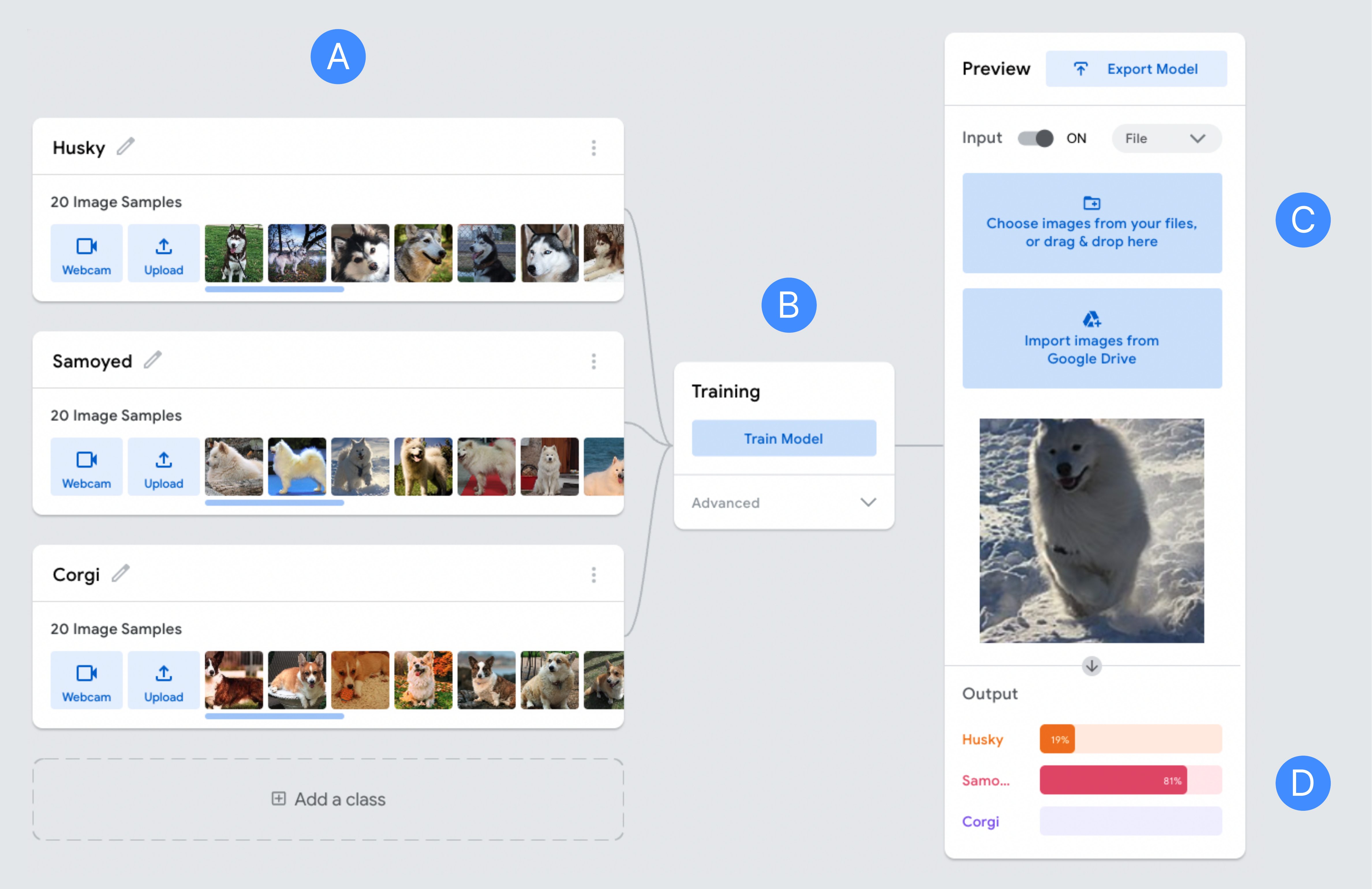}
    \caption{The Teachable Machine interface. \textbf{A}: class modules where users can drag-and-drop image files to upload training data. \textbf{B}: training module with a button for initiating model training. Training typically takes less than 30 seconds. \textbf{C}: input module where users can upload an image for the model to evaluate. \textbf{D}: output module with the model's class probability scores on the evaluated image.}
    \label{f:tm-interface}
    \Description{The Teachable Machine interface against a gray background with white modules. 3 modules on the left represent classes (Husky, Samoyed, and Corgi) and contain 20 training images each. A training module in the middle features a blue  Train Model button. A module on the right allows users to upload evaluation images via Google Drive or local files. Here, an image of a samoyed running against a snowy backdrop is evaluated. The output features a bar chart visualization features detected probabilities for each class: 19\% for Husky, 81\% for Samoyed, and 0\% for Corgi.}
\end{figure*}

\subsubsection{Activity}
\label{s:activity}
We structured our activity based on the industry-standard double diamond design process \cite{design-council-2005}. Yang et al. \cite{yang2020difficult} showed that AI design challenges persisted throughout the double diamond design process. Since we focused specifically on communication of the design material (the last part of the second diamond) and did not want to subject participants to excessive fatigue, we designed an abridged version of the design process where we targeted activities typically associated with the second diamond (solution exploration and presentation). We did so by providing participants with resources---such as user research insights and personas---that they may otherwise need to dedicate substantial time and resources to set up (as part of the first diamond). This way, we ensure they have ample time and support to thoughtfully craft their solution.

The following design prompt was given to participants at the beginning of the activity:
\begin{quote}\textit{Your company likes to invest in new ideas, particularly ones that use machine learning. You and another designer have an idea for a mobile app that uses machine learning to help users understand their eating habits. The basic idea is a photographic food journal to help users understand whether they are eating well. [...] You and your partner developed a preliminary persona to help you both stay focused on a potential user. You need to design and present a proof-of-concept for the app.}
\end{quote}

This prompt served two purposes: to provide participants with a concrete starting point and to focus their attention on UX challenges associated with the AI aspects of the app. While food and diet tracking are widely understood concepts and need no in-depth introduction, we acknowledge that there may be gendered biases around this topic. To mitigate this, we created both a female and male persona---keeping characteristics between them constant except for their photos, names, and background information. We counterbalanced the personas by randomly distributing the female persona to 14 participants and the male persona to 13. 

To obtain a better understanding of AI as a design material, participants trained image classification models in Teachable Machine as part of the design activity. We provided participants with all training and evaluation images, which were in the form of 3 datasets. The datasets all contained the same 300 images, which we randomly sampled from the EPFL Food-11 dataset \cite{food-dataset} of 16,643 food images. We chose to sample\footnote{When sampling images from Food-11, we preserved the ratio of the number of images between classes to mimic the imbalance in the original dataset.} from Food-11 due to its size and class diversity. Although the 3 datasets participants received contained the same images, they were labelled in distinct ways: one had 2 classes, one had 3, and the other had 5. We organized each set of classes based on a specific mental model: 2-class was a representation of the healthy-unhealthy binary, 3-class was based on how restaurants categorize food on menus (appetizers, entrees, and desserts), and 5-class aligned with food groups in the MyPyramid food pyramid \cite{britten2006development} published by the USDA (dairy, fruits, grains, protein, and vegetables). Participants were encouraged to train all 3 models and select one to incorporate into their solution. We also provided participants with the latest version of MyPyramid alongside the datasets in case they were unfamiliar with it.


    
    
    
    

Lastly, participants received guidelines for creating their design presentation, which they would use to communicate their solution to stakeholders on their team as if they were taking part in an interdisciplinary design critique. We left the contents and medium of this proposal up to the participant; the only requirements we set were that it should be in the form of approximately 10 slides/pages, and that it could be exported as a PDF. 

All materials used in the study (design prompt, user research insights, persona, datasets guide, MyPyramid, and presentation guidelines) were packaged into the folder participants downloaded at the beginning of the session. The PDF of our materials packet is available in our Supplementary Materials. After debriefing participants about the materials, we gave them time to work, checking in occasionally to answer questions and give notice of remaining time. The activity ended when the participant finished the design presentation and shared the presentation PDF with the research team member via email. This work period was typically 75 minutes in length.

\subsubsection{Interview}
We conducted semi-structured interviews with our participants upon the conclusion of the activity. The discussion centered around two main topics: Teachable Machine and the presentation. We asked participants how they used Teachable Machine throughout the design process and whether it enabled them to achieve the goals they envisioned for their solution. We also discussed how the tool can be incorporated into demos and presentations, as well as general usability. We reviewed the presentations and asked about concepts and aspects of the process they thought were most important to communicate to stakeholders, in addition to certain design decisions they decided to show on their slides. Our full interview protocol is available in our Supplementary Materials. Our interviews were typically around 30 minutes in length.

\subsection{Data Analysis}
\subsubsection{Analyzing Interview Transcripts}
Our interviews were recorded using Zoom and thus came with auto-transcriptions. The first author manually reviewed the recordings alongside the generated transcripts and corrected any mistranscribed words. Two authors took a first pass over the data and identified salient regions of the transcript that may be of interest for further analysis. The first author then performed open and axial coding on those regions to identify themes and establish connections across themes. The themes were partially informed by frameworks in previous literature on AI user experience design (see Section \ref{s:rr-ml-material}). We paid especially close attention to how UXPs' experiences in this study and more generally in their work may impact collaborative practices with stakeholders they work with. 

\subsubsection{Analyzing Design Presentations}
The research team derived a 5-step visual coding process to collaboratively code the design presentations created by participants. We imported all presentations into a FigJam\footnote{FigJam is a collaborative whiteboarding tool: https://www.figma.com/figjam/} canvas and wrote virtual sticky notes on top of relevant areas in the presentation slides to denote a code. For visually dense areas in the slides, such as a wireframe, we drew boxes around areas of interest and labelled each box with one or more sticky notes. Our process is further described as follows:
\begin{itemize}
    \item \textbf{Pass 1 (individual):} We divided up the 27 presentations among 4 members of the research team and wrote sticky notes to features of interest. This is akin to open coding. We then reviewed others' open coding and added codes as we saw fit. 
    \item \textbf{Pass 2 (individual):} Each coder then grouped their codes from Pass 1 into high-level codes. Here, we also employ a more formal feedback-sharing practice where each coder also wrote brief reviews of others' coding. The reviews consisted of comments indicating what they thought was interesting or thought-provoking and how it connected to their own coding.
    \item \textbf{Pass 3 (group):} The research team came together to discuss common themes across high-level codes. This is akin to axial coding. We drafted a codebook with these themes and recoded where necessary. We also wrote summary memos based on themes in our codebook. 
    \item \textbf{Pass 4 (group):} The team came together and resolved confusion emerging from the codebook. We modified some codes to clarify our analysis and minimize confusion. We also revised our memos accordingly.
    \item \textbf{Pass 5 (one person):} The lead researcher then consolidated any leftover inconsistent codes in the FigJam canvas and summarized memos across all coders.
\end{itemize}

The memo summaries were then combined with insights from our interviews to form our findings. Note that we did not calculate any measures of coder agreement, such as inter-rater reliability, as the purpose of our coding was not agreement, but to discover emergent themes \cite{mcdonald2019irr}. Our 7 themes in our final codebook and their descriptions are summarized in Table \ref{t:themes}.

\begin{table*}[h]
\centering
    \begin{tabular}{p{5cm} p{9cm}}
    \toprule 
    Theme & Description\\
    \midrule 
    Current challenges & Communication and collaboration challenges participants encountered in their professional practice when designing for AI. \\
    
    Communication to stakeholders & Methods and techniques that communicate the proof of concept solution to theoretical stakeholders participants are working with. \\

    Presence of AI/ML & Indication of presence of AI, or lack of its presence, or an interpretation of the implications of its presence. \\

    AI/ML-User Interactions & Design affordances that support user interaction with ML model and related components. \\

    AI/ML Framing & Descriptive framing of AI/ML's goals and specifications in the context of the solution.\\

    Concerns/Risks & General concerns and/or ethical issues that may stem from app. \\

    AI/ML Next Steps & Future plans, actionable by participant and/or their team, for AI/ML components in the app if they were to continue development beyond the study. \\
    
    \bottomrule
    \end{tabular}
    \caption{Our 7 themes that emerged from analyzing participants' design presentations and post-activity interviews.}
    \label{t:themes}
    \Description{A 2-column table with headings Theme and Description. There are 7 rows with the 7 high-level themes alongside their descriptions.}
\end{table*}

%% file: sections/results.tex
By analyzing the design presentations and interview transcripts from UXPs during this study, we identified emergent themes in how UXPs work with and communicate AI as a design material. These themes include difficulty of collaboration due to domain gaps, emphasis of AI model accuracy, change in collaborative strategies upon exposure to interactive AI, awareness of both AI benefits and risks, and formulation of actionable next steps. Below, we present these findings in detail.

\subsection{Domain Gaps Induced Collaborative Friction}

We asked UXPs who had prior experience working on AI-enabled interfaces about their collaboration experiences with non-UX stakeholders. Their comments revealed many challenges to communicating and collaborating effectively with AI engineering teams, which were grouped as an emergent theme of ``Current challenges.'' We found that despite calls for co-creation processes between UX and AI teams \cite{subramonyam2021towards}, work in the two domains still occurred independently and linearly in practice. In particular, P14 aptly summarized a common trend when UXPs work with AI teams, which was also shared by many other UXPs: 

\begin{quote}
\textit{``There is this dedicated data science/AI team that tries to create that algorithm or gather data around it. After that, the design team and in involved and how [we] want to surface it and allow the users to interact with it''} (P14).
\end{quote}

P8 considered this a \textit{``very top down''} approach in the sense that \textit{``the engineering team [says], here's all the work we already did, take it and make [the UX] better,''} and wished for user-facing work to begin earlier in the process. P4 observed that early involvement is possible if the technology has not been defined yet. In those cases, P4 preferred to \textit{``first start with user and market research, because developing technology is very costly and engineers' time is very precious.''} However, this approach requires UXPs to be aware of AI-based product planning discussions and to have a seat at the table when discussions occur, which was remarkably rare as only a couple UXPs recalled being a part of them. 

The gap between the two domains was not only viewed as an obstacle to effective collaboration, but also eroded UXPs' trust in AI teams. P10 summarized design as \textit{``always a balance of engineering cost and user experience''} and therefore, if \textit{``I know nothing about [the technology], it will be really hard for me to establish a collaborative or efficient workflow with my engineers.''} This gap was so prominent in many cases that UXPs were unsure if what they designed was actually using AI/ML: \textit{``it was being developed by our machine learning division, so in that sense, I think it's AI-powered, but I'm not sure what was done''} (P4). Perhaps it was due to this opaqueness that UXPs began questioning the value of work produced by AI teams. P5 expressed their frustration and loss of trust as follows: 

\begin{quote}
    \textit{``The teams I work with are so called `machine learning' teams, but to be honest, looking at our current site experience, I have zero faith in what this team does, or their capability of using machine learning to deliver something relevant. [They] are supposed to learn [user] preferences and eventually  get our model to be smarter, but I never really see that happen. So I always question what's the purpose of the machine learning team''} (P5).
\end{quote}

In response to the domain gap and its consequences on collaboration, UXPs have actively attempted to create better ``bridges'' with AI stakeholders, but with limited success. P1 describes this awkward process as \textit{``trying to bridge this gap where we're trying to speak the same language but really we have very different skill sets.''} P8 adds that poor timing was also a factor: 

\begin{quote}
    \textit{``[The AI teams] were so detached from users and so focused on the data that I really did have to bridge the two [domains], and I still feel like I didn't do a great job because I just didn't know the data that well, or what was happening, until after the product ships, which is way too late''} (P8).
\end{quote}

Given these prior concerns, we postulate that UXPs' experiences with hands-on AI experimentation using Teachable Machine and the content they chose to highlight in their design presentations may provide valuable clues on how such collaborative friction may be mitigated. We further explore this in the following subsections.

\subsection{Accuracy was Considered Important but UXPs Struggled to Communicate It}
\label{s:accuracy}
\begin{figure*}[h]
    \centering
    \includegraphics[width=0.9\textwidth]{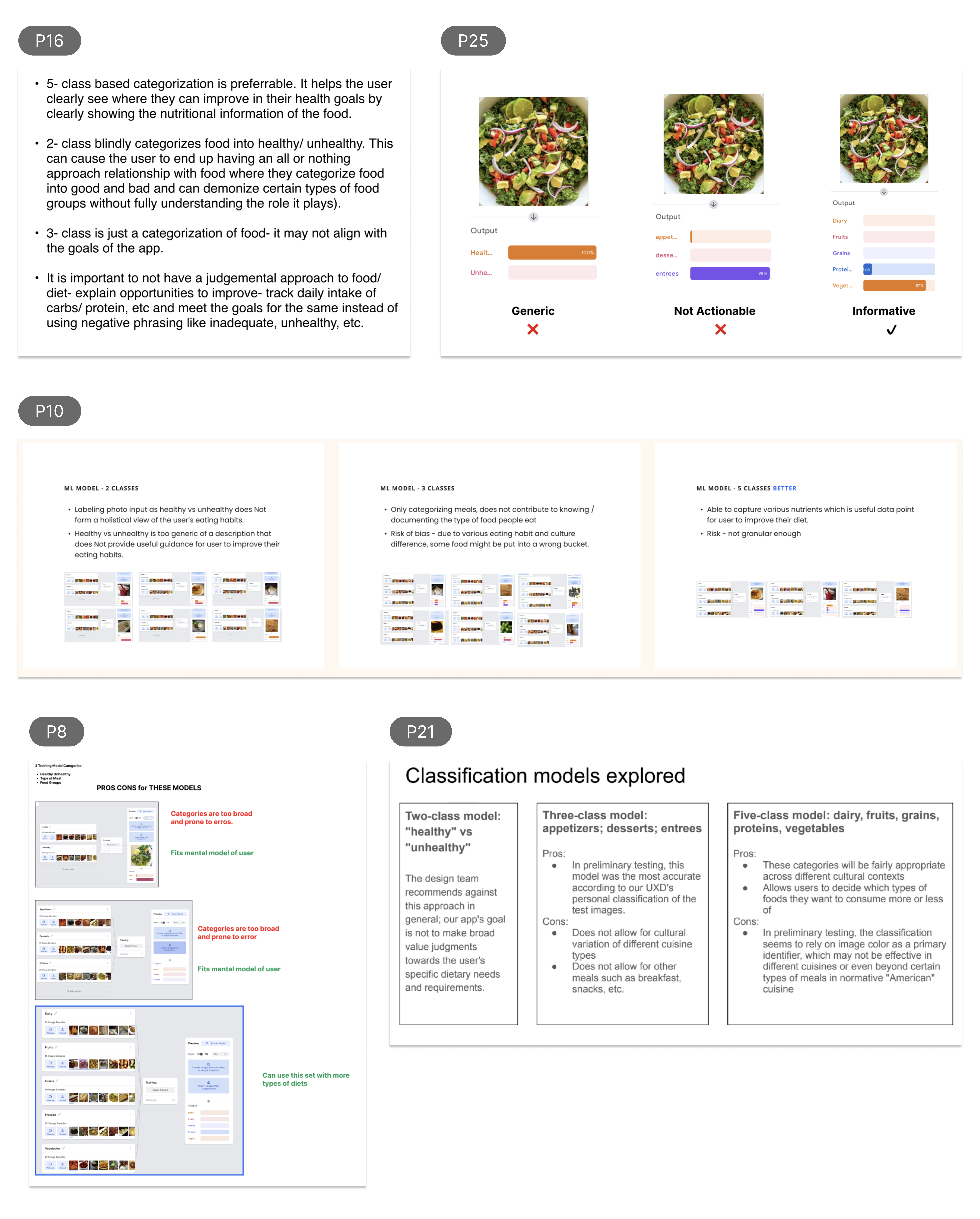}
    \caption{Examples of how UXPs communicated model selection rationale in their presentations. Some dedicated one slide to their explanation, while P10 dedicated several.}
    \label{f:rationales}
    \Description{5 examples of how UXPs compared different models in their design presentations. Some used only text and bullet points to justify their decisions, while others used a combination of text and screenshots of the different models from Teachable Machine.}
\end{figure*}

When asked about key information to communicate to stakeholders in their design presentations, UXPs brought up model selection rationale, customer value, engineering costs, and most notably, model performance. When asked about the comparison of models in their presentation (see Section \ref{s:activity} for descriptions of the different models), P25 stated that \textit{``I always pay attention to rationale, so in my presentation I make explicit why I chose this specific model versus the other two.''} P21 echoed that they provided a brief introduction of the three models and \textit{``what I see as the pros and cons of each of them.''} We showcase how P21 and P25 communicated their rationale for model selection in their design presentations, along with other UXPs who did so, in Fig. \ref{f:rationales}. P2 and P5 both stated that they prioritized helping stakeholders \textit{``understand what's the benefit to customers and how [the solution] will benefit the business''} through competitive analysis of other food tracking apps. In addition to customer and business value, P10 also mentioned that \textit{``definitely engineering costs will be something that leadership will pay attention to.''}

By far the most commonly mentioned aspect to communicate was model performance---more specifically, accuracy. As P17 put it, \textit{``the most important thing would be coming to stakeholders with accuracy, because if it's not accurate, then the rest is meaningless.''} Indeed, UXPs placed great emphasis on accuracy and some incorporated it into their design goals from the very beginning: \textit{``one of like the design goals that I had for the app was that it gives accurate and reliable information''} (P19). P4 illustrated a reason for why they, and possibly others, designed with accuracy in mind: \textit{``my assumption would be that if it's more accurate it's more likely to meet the users needs.''} Presenting accuracy was not only seen as informative, but also a means of generating discussion to identify steps for improvement. P14 anticipated that discussions of accuracy will arise in stakeholder meetings to identify problem sources such as data quality:

\begin{quote}
    \textit{``During the stakeholders meeting this will come up too, because if you are like, oh this is not accurate enough, like what is the problem? Is it that we don't have enough data or is it the quality of the data?''} (P14).
\end{quote}

Even if the accuracy was low, P5 still considered it important to communicate as it helps \textit{``figure out a way to make it more accurate in the future, as we launch this product.''} Similarly, P19 saw the presentation of accuracy as a way to begin discussions on the model and data and work toward ``closing the gaps'' in necessary information to build a robust product:

\begin{quote}
    \textit{``I think PMs would be really interested in knowing reliability or accuracy of information [from the model]. So if there's a way of being like okay, I trained on these three classes and the three classes are based off of X number of samples or like X volume of data, and these are things that we are working towards to closing the gaps on''} (P19).
\end{quote}

\begin{figure*}[h]
    \centering
    \includegraphics[width=1\textwidth]{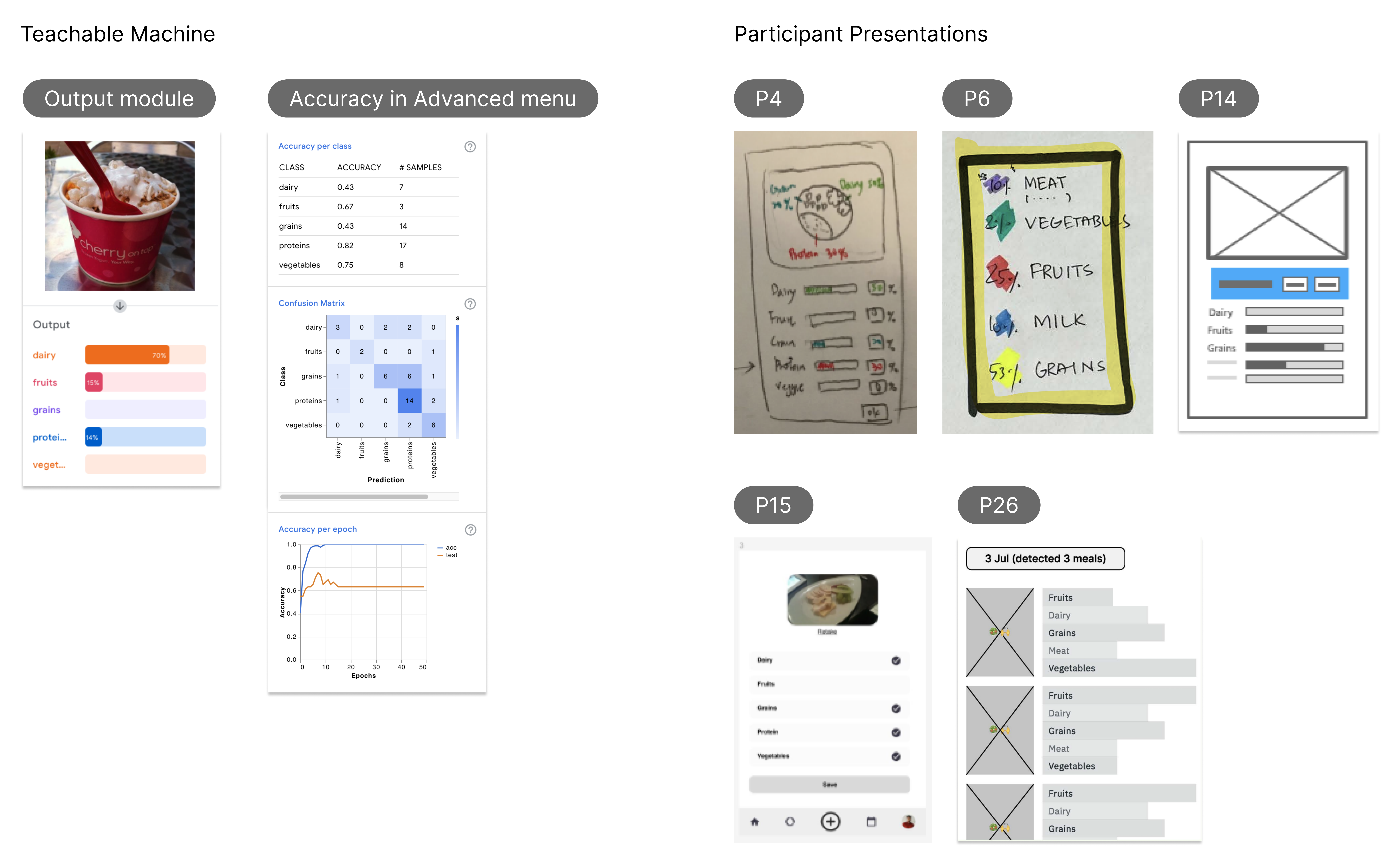}
    \caption{Output module and accuracy evaluation options in Teachable Machine (left), alongside examples of evaluations shown on UXPs' slides (right). Deviations from Teachable Machine's design include the absence of bars when expressing percentages (P6) and using a binary yes/no to indicate the presence of a class rather than a percentage (P15).}
    \label{f:evaluations}
    \Description{Teachable machine produces a bar chart visualization of class probabilities for every image evaluated. An accuracy table, confusion matrix, and accuracy-loss graph can also be accessed in the Advanced menu. Those are shown on the left while examples of how UXPs displayed evaluations in their presentations (which closely resemble teachable machine's bar charts) are shown on the right.}
\end{figure*}

Given this emphasis on accuracy, we expected UXPs to derive a variety of reporting strategies to communicate accuracy to stakeholders. However, upon analyzing their design presentations and specifically the emergent theme of ``Presence of AI/ML,'' we failed to find a single presentation that reported accuracy directly. Reviewing the communication of AI/ML results to users in the ``AI/ML-User Interactions'' theme, we also found the style of reporting to be relatively homogeneous. Participants primarily used the bar chart visualization in Teachable Machine's output module (see Fig. \ref{f:tm-interface}D), with slight modifications, in their presentations. A key observation here is that Teachable Machine's output module shows \textit{class probability scores} on a particular image being evaluated, \textit{not model accuracy}. Teachable Machine offers an option to generate per-class accuracy tables and confusion matrices for trained models, but it can only be accessed in an ``Advanced'' menu under the Train button in the interface (see Fig. \ref{f:tm-interface}B). Several UXPs accessed this menu but told us that they did not understand its contents and avoided it as a result. In Fig. \ref{f:evaluations}, we gathered a collection of attempts to show model accuracy from UXPs who mentioned the importance of communicating accuracy in our interviews, juxtaposing them with Teachable Machine's output module.

These findings may lead one to ask: why do UXPs place so much emphasis on model accuracy (or perceived accuracy, as they did not conduct formal evaluations) when many AI products in the real world deliver user value with only moderate model performance \cite{ai-errors}? We posit that this may be due to Teachable Machine's prominent display of quantitative metrics on evaluated images, as well as unfamiliarity with how to handle and guide the user through model uncertainty. Any accuracy metric not in the proximity of 100\% may appear to run counter to UXPs' goals of ensuring a safe and reliable experience for users. Given inevitable inaccuracies in AI models, we recommend future tools and process models to support UXPs in making sense of performance metrics and designing for uncertainty.


\subsection{Tinkering with AI Shows Promise in Improving Communication with Stakeholders}
\label{s:findings-tinkering}
Regardless of whether they were able to clearly communicate accuracy, UXPs' comments and the emergent theme of ``Communication to Stakeholders'' in their design presentations suggest UXPs found hands-on experimentation with Teachable Machine valuable in preparation for communicating with their stakeholders. The interactive nature of the tool presented an engaging way to communicate AI concepts to executives, as \textit{``usually executives have a hard time imagining what this thing could look like if you don't show a visual''} (P5). Some said that showing a simple demo in Teachable Machine, even if it is nowhere near the model that will eventually be used in the actual product, is a \textit{``good start to getting our knowledge on the same grounds''} (P13) and \textit{``could help to sell the idea and make it feel less scary''} (P1). UXPs also appreciated the ability to independently test their ideas in Teachable Machine before presenting them to others. P12 specifically mentioned that it can prompt discussions with engineers: \textit{``maybe I can [build a model] myself, test it, and give some comments to my engineers and we can discuss together.''} UXPs were careful to note the limits of the tool: while it provided a starting point for conversation, the model trained with it was not something they would want to put out into the real world. As P23 put it, \textit{``I don't want to be a data scientist, like I won't want to have to train [the production-level model] or edit it.''} Thus, much of the value in tinkering with AI in a tool such as Teachable Machine is that it broadens common ground for communicating with engineers and articulating the value of the technology (or lack thereof) to executives.

UXPs primarily responded with one strategy when asked how exactly they would show the results of their tinkering with stakeholders: walking through a live demo in Teachable Machine. The demo, according to many, should include all the steps they personally completed in the design activity---data uploading, training, and evaluation. With a bigger team, P18 suggested a \textit{``workshop that everyone tries it all together.''} This UXP-led demo is made possible due to the simplicity of the Teachable Machine interface; P15 said that they would see themselves \textit{``taking [the tool] right now and showing it to my product manager [since] the UI is simple enough.''} P8 added that \textit{``I could use this and explain [machine learning] to my mom.''} Besides live demos, UXPs also had suggestions for features this tool could add to improve the experience of sharing results. P7 was interested in exportable tutorials so that the demos do not necessarily need to happen synchronously. Relatedly, P20 wished for a more \textit{``consumable deliverable, a thing that I could quickly plop in a PowerPoint deck and show it to someone.''} Indeed, the desire for integration with other tools was also echoed by P3 and P16. P16 said they wanted Teachable Machine to be integrated within their native design environment because they felt like they \textit{``couldn't navigate between two different tools''}, while P3 wanted to bring their other files (and discussions) into Teachable Machine: \textit{``if it can allow me to attach [other] files and loop on another discussion thread, that'd be cool.''} A few participants (P11, P25, P26) also pointed out that explaining key terminology used throughout the tool, even basic words such as ``model'' or ``classes,'' would create a smoother onboarding experience for those who are brand-new to AI.

While we observed improvements in UXPs' high-level understanding of AI after tinkering with Teachable Machine, we realized, upon inspecting their design presentations, that the experience did not necessarily succeed in calibrating expectations of AI's capabilities. A few UXPs overestimated model capabilities and assumed they could also extract information on food weight based on the given dataset. We consider this an overestimation since the data for this supervised learning task did not contain any weight information in its labels. UXPs may have done this as an attempt to cover up deficiencies in the design material---they recognized that a major limitation of applying AI to food tracking was that it fails to account for portion size. While assuming a model is more capable than its true abilities is undesirable in many cases, we actually see it as valuable in the context of design communication. Stakeholders with expertise in AI may ask further questions about the necessity of this feature and engage UXPs in discussions about feasibility, promoting more effective communication.

\begin{figure*}[h]
    \centering
    \includegraphics[width=1\textwidth]{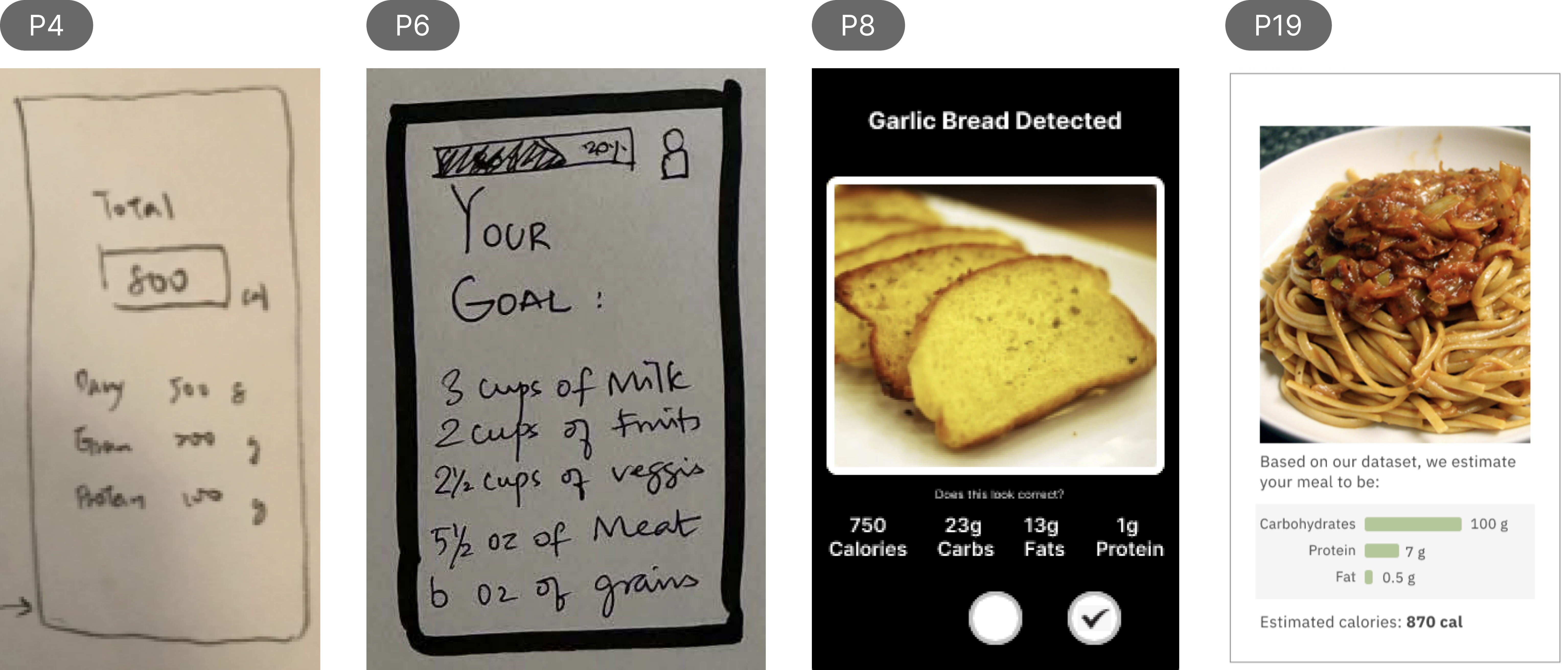}
    \caption{Snippets of UXPs' slides where they incorporated weight information from model output into their designs.}
    \label{f:assumptions}
    \Description{4 examples of how UXPs made capability assumptions about the model in their design presentations. These include inferring food weight and specific nutrients (e.g., calories and carbohydrates).}
\end{figure*}

\subsection{UXPs Paid Attention to Both Benefits and Concerns of ML}
\begin{figure*}[h]
    \centering
    \includegraphics[width=1\textwidth]{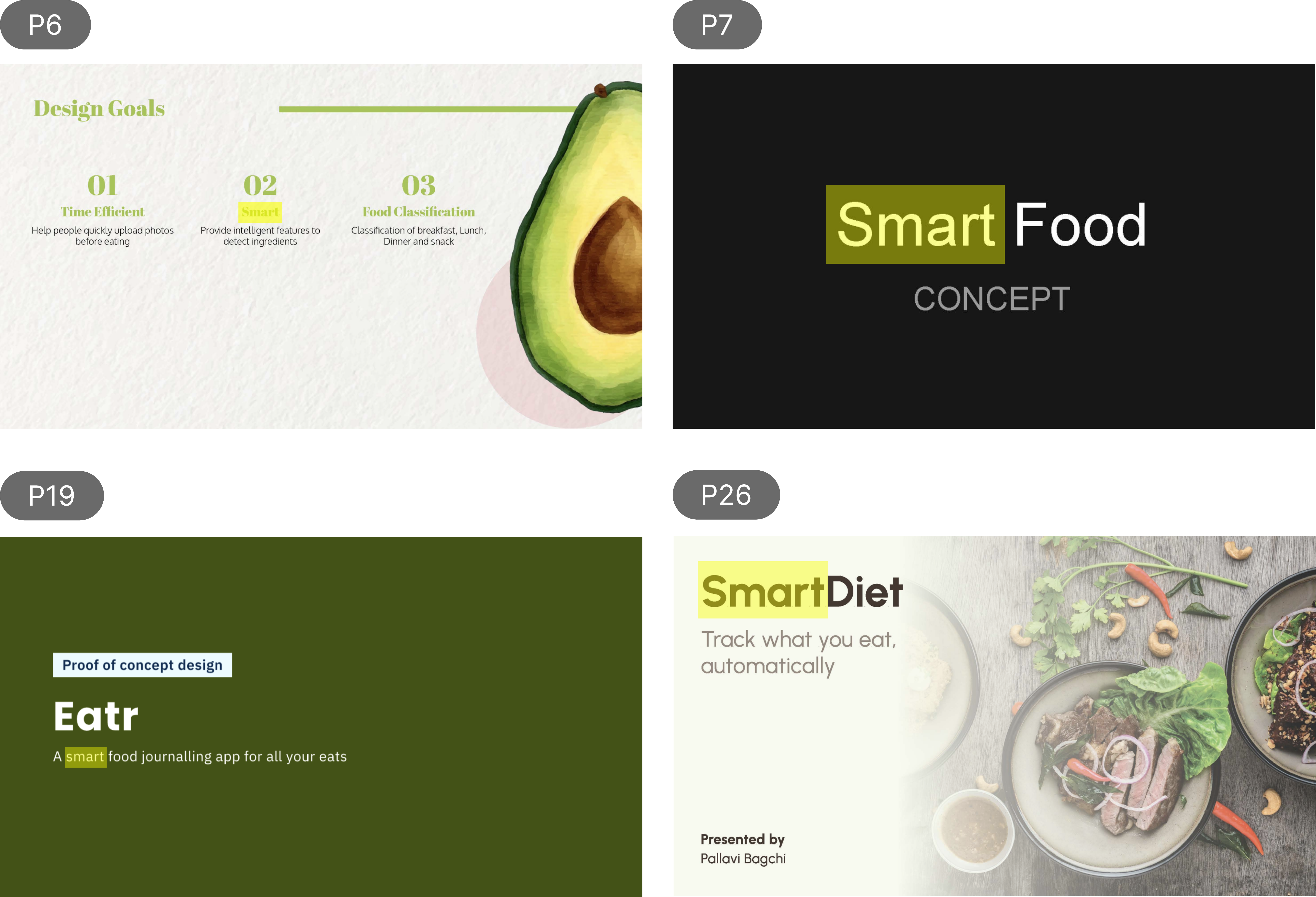}
    \caption{UXPs' use of the word ``smart'' in their title and design goals slides. Highlights were added by the research team for visibility.}
    \label{f:smart}
    \Description{4 examples showing how UXPs employed the word ``smart'' in their design presentations. Some had it on the title as their design's name while others described it in their design goals.}
\end{figure*}

As we reviewed participants' design presentations, we paid close attention to how they framed AI in the context of the design exercise, both in terms of the benefits AI can bring as well as concerns with an AI-based approach. Within the common theme of ``AI/ML Framing'', many UXPs agreed that AI can add significant value to a food tracking app with its ability to make the process \textbf{``smart,''} \textbf{``fast,''} and \textbf{``easy.''} By \textbf{``smart,''} we believe they were referring to the ability of AI to make relevant recommendations based on existing user data. For example, P16 discussed how AI-enabled recommendations can improve user experience over more general, rule-based approaches: \textit{``instead of having an entire screen describing the general `rules' to eating healthy, try to throw in a [personalized] health tip like `it is recommended that \_\_ g of protein must be consumed every day.'''}  P7 summarized their solution as being able to \textit{``provide nutritional feedback on users' eating habits.''} We collected some examples of UXPs framing their solution as ``smart'' and showcased them in Figure \ref{f:smart}. UXPs also identified excessive manual effort required in current diet-tracking apps as a pain point. Thus, they strove to design a solution that was \textbf{``fast''} and \textbf{``easy,''} and framed AI as being able to save time and effort in the user flow. This can be done by leveraging users' established habits of taking pictures of their food before eating (P3) and automating repetitive tasks such as manually inputting ingredients (P18). Furthermore, the novelty of AI also served as a motivating reason to use it: P2 wrote on their slides that the idea of applying AI to diet planning was a novel and innovative one, and therefore has positive business impact.

UXPs also shared numerous risks they identified with using AI in their solution. They communicated their concerns in their design presentations, and many dedicated a specific slide to listing identified risks. This was prevalent across participants and organized under a theme of ``Concerns/Risks.'' Concerns included potential issues with both the properties of AI itself, as well broader cultural issues with applying AI to the domain of food tracking. Notable examples of the former include poor accuracy, lack of explainability and consequent loss of user trust, and lack of input from medical experts. In particular, P3 was worried that the lack of accuracy can hinder the user experience as the manual work of correcting the errors may surpass the benefits of automation. Examples of cultural concerns primarily centered around bias and unintended societal implications. These included cultural biases induced by the training data, biases in the definitions of some model classes (for example, P1 pointed out that ``(un)healthy'' can vary person-to-person and culture-to-culture), lack of transparency about privacy policies when training on personal food data, and over-judgements of eating habits, which may lead to food-shaming and health risks of under- or over-overeating. Along these lines, P2 pointed out that relying on habit-forming behaviours (which was previously highlighted as a benefit of using AI) can amplify bad eating habits, which may be difficult to correct and raise the need for medical attention. P7, who introduced a ``healthiness score'' metric based on the AI model's outputs, admitted that there can be some gamification to reach a high score, which might not be in the users' best health interests. 

Overall, we were pleasantly surprised to see that after some brief (< 2 hours) tinkering with a simple tool like Teachable Machine, they were able to realize and communicate many AI-related risks in their solution. This is especially true since most participants had no prior experience designing with AI. 

\subsection{UXPs Proposed Actionable UX and AI Next Steps}
\label{s:next-steps}
Since the context of the study was crafting an early proof-of-concept within a team-based setting, it was important for us to observe how UXPs communicated next steps to their team. We found that UXPs were able to identify a series of concrete next steps in both UX and AI domains, evidenced by their comments and the ``AI/ML Next Steps'' theme in their design presentations. 

From the UX perspective, many expressed the desire to conduct more user research as well as test some of their current designs with users to solicit feedback. P23 specifically mentioned the role of research in understanding user goals and selecting a model to use: 

\begin{quote}
    \textit{``I also would've done more foundational research with potential users to discover what the best starting point categories would have been. I don't think I would have started with a universally applicable classification, like what does healthy or unhealthy mean for somebody training for a marathon?''} (P23). 
\end{quote}

Additionally, P22 shared that they worked at a large organization with experts specializing in best practices and guidelines when incorporating AI into user-facing products, and said that they would \textit{``probably go to them with this concept and be like, hey, what are some problems that you can see that we should address?''} P13 bought up the issue of positionality and the need to more carefully consider how their solution addresses it: \textit{``how do we position ourselves within the super personalized idea of food? And [set some boundaries on] what we definitely should not do and where we should not go.''} Many also admitted that due to the study's time constraints, they did not have time to mock up as many mid- or high-fidelity screens as they would have liked. P27 mentioned that they would be interested in exploring another user flow in addition to designing higher-fidelity interactions.

From the AI side, some mentioned that they would like to dedicate more time to learning about different AI models and how they can better fit the solution. P10 felt that researching AI \textit{``will be the most time consuming part just because I don't know too much in this area (AI).''} Outside of models, UXPs also expressed desires to learn more about AI in general to identify further opportunities for UX: \textit{``I would definitely want to go through more resources for machine learning and any other research about how machine learning can be applied to UX or vice versa''} (P18). A few UXPs were also interested in connecting AI outputs to other resources, such as nutritional information from health experts, to provide better tips and recommendations for healthy eating. Those who raised AI-related risks and concerns also brainstormed ways to address them going forward. P1 wrote in their presentation (and also confirmed in the interview) that the team should \textit{``ensure that the model is trained with foods from as many cultures as possible''} to reduce cultural bias. P15 also echoed this sentiment, but suggested that the models should also be trained with photos \textit{``from may environments to account for variability in angle, lighting, etc.''}. P18, among many others, saw user feedback as a path to improving model performance and suggested designing affordances for users to submit feedback by confirming or correcting model outputs. 

The willingness of UXPs to get more involved in AI-adjacent work, as well as their thoughtful reasoning of next steps, is a testament to the benefits of their early involvement in the AI design process. As many UXPs have told us, their entry point into the process occurs when they are asked to build a UX layer on top of a completed model. We see from their formulation and communication of next steps that UXPs could be effective collaborators who work alongside or even before AI teams, rather than after.

%% file: sections/discussion.tex
Our findings revealed new opportunities for concepts and tools to facilitate work at the intersection of UX and AI. We first discuss applying the concept of prototype fidelity to AI models as a sensitizing concept for UXPs in interdisciplinary teams. We then present an opportunity for AutoML tools to help teams design and build AI systems collaboratively. Lastly, we propose probabilistic user flows to help bridge the gap between the different metrics for success used by UXPs and technical stakeholders.

\subsection{AI Model Fidelity as a Sensitizing Concept in UX}
In UX, practitioners use the term \textit{fidelity} to refer to a prototype's resemblance to a completed final product \cite{walker2002high}. A low-fidelity prototype has very limited functionality and is meant to be a limited-effort artifact to depict early concepts and layouts, whereas a high-fidelity prototype faithfully represents the possible interactions in the eventual interface but trades off accuracy for speed of creation \cite{rudd1996prototype}. Working in incremental iterations from low to high fidelity is critical in UX practice to test design assumptions, ensure the inclusion of user feedback, and identify usability flaws before they are cemented into the interface \cite{camburn2017design}. 

In our study, we observed participants rapidly train, iterate on, and communicate multiple AI models. This mode of experimentation precisely captures the spirit of low-fidelity prototyping. They communicated their model explorations in their design presentations (see Fig. \ref{f:rationales}) and reasoned about how the models can offer additional capabilities and address limitations in future iterations (see Section \ref{s:next-steps}). We can think about these future models as ``higher fidelity'' than the ones trained during the study: they likely require more data, time, and AI expertise to train, but they are a closer representation of what a user may experience in a real product. 

We therefore extend the concept of fidelity to AI models as a \textit{sensitizing concept} to support UXPs' communication of AI as a design material in interdisciplinary teams. 
As an example of how this may scaffold information sharing with collaborators, let us consider the key characteristics of a low-fidelity image classification model. First, the model should serve as a simple starting point. A low-fidelity model is not meant to demonstrate the full capabilities or achieve comparable performance levels as a deployment-ready model, but rather to determine which target classes will be valuable to the user. These classes can be communicated as a data requirement to AI practitioners to guide data preparation. Second, the model should be easily modified and iterated upon. This way, a new prototype can be created in response to user feedback or new user research. This iteration, which may shift overall product direction, can be relayed to product managers and business stakeholders. Third, the model should not require significant effort to use and test. In collaboration with AI practitioners, UXPs may gather some sample data and use a tool like Teachable Machine to train experimental models and validate UX hypotheses before model specifications are finalized. On the contrary, high-fidelity AI models should be robust in functionality and performance, and thus require the expertise of AI practitioners to train, but compromise its ability to iterate quickly as a result. 

More generally, we see AI model fidelity as a \textit{technique-agnostic} sensitizing concept that can be applied to models beyond the simple classifiers explored in this study. Let us consider the case of generative AI. The barriers to tinkering with large, pre-trained generative AI models have been substantially lowered due to tools such as the OpenAI playground and ChatGPT, but the challenges of communicating AI remain---the models remain black-boxed despite having grown significantly more complex. As UXPs tinker with default base models to gain insight into their behaviour and compare capabilities across different base models, they may designate them as low-fidelity models. They are easily usable off-the-shelf, but are a general-purpose solution that comes with hidden biases and risks \cite{weidinger2022taxonomy}. UXPs can then iterate towards a higher degree of fidelity by prioritizing user needs and working with relevant stakeholders to tailor base models to domain-specific use cases reflective of those needs via model fine-tuning. Regardless of the particular technique used to implement AI, whether it be regression or pre-trained transformers, the importance of clear, effective communication of AI as a design material will thus remain.

\begin{figure*}[h]
    \centering
    \includegraphics[width=1\textwidth]{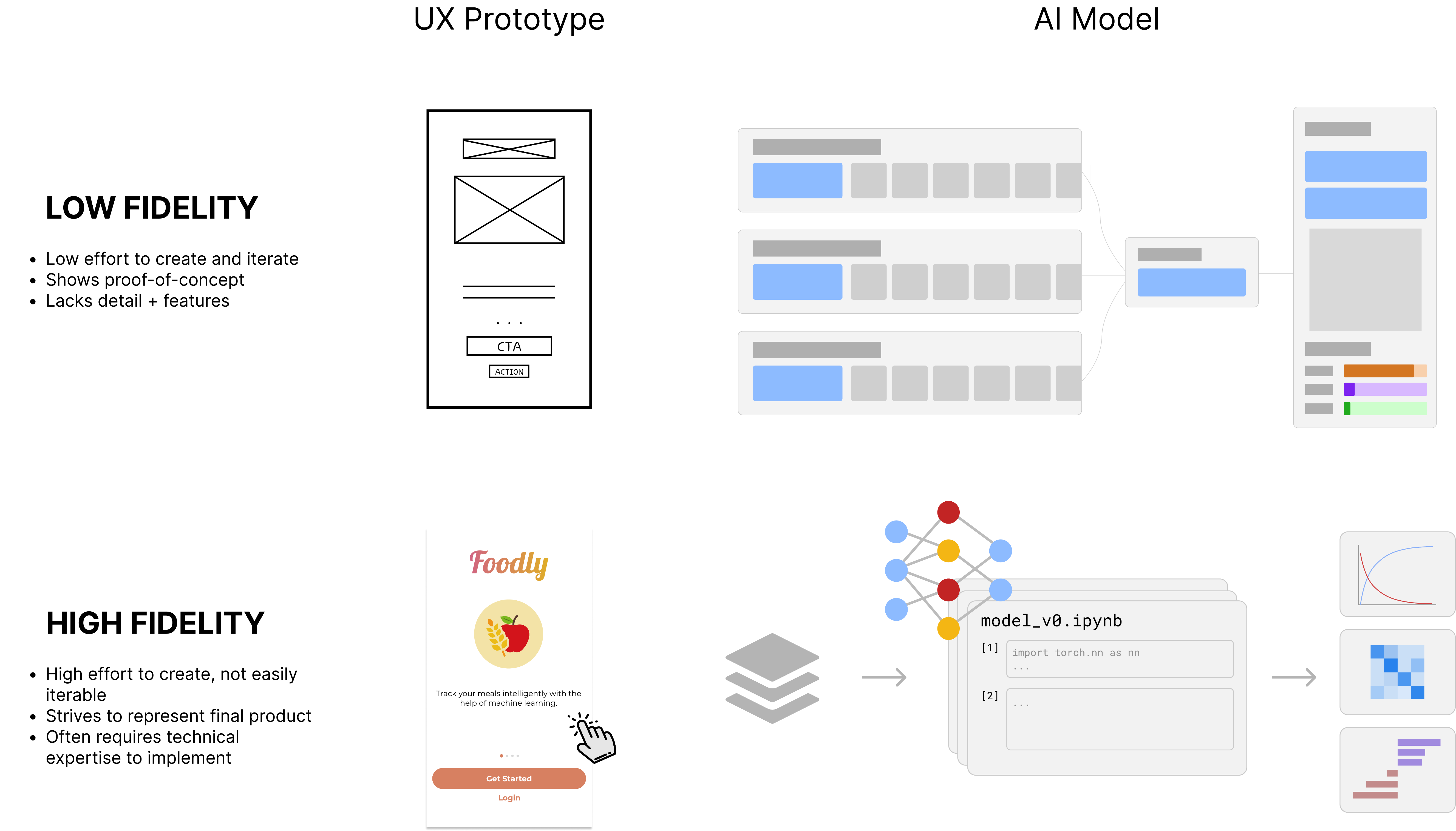}
    \caption{Low- and high-fidelity representations of UX prototypes compared with those of AI models.}
    \label{f:fidelities}
    \Description{A two by two diagram with low and high fidelity as rows and UX prototypes and AI models on the columns. A low-fidelity UX prototype is a simple black and white wireframe, whereas the high-fidelity prototype has well-articulated fonts, logos, buttons, and colours, along with a hand icon indicating interactivity. A low-fidelity AI model is similar to one created in Teachable Machine, while a high-fidelity model takes in a dataset as input, contains model code in a Python notebook, and outputs 3 visualizations used for performance evaluation: accuracy-loss chart, confusion matrix, and feature relevance bar plot.}
\end{figure*}

We illustrate low- and high-fidelity AI models in Fig. \ref{f:fidelities} alongside low- and high-fidelity interface prototypes. Just like how iterating across fidelity levels in prototyping can stimulate and enrich between UXPs and non-UXPs, the same may occur when UXPs leverage this sensitizing concept in designing AI-enabled interfaces. 

\subsection{AutoML as a Boundary Object}
Hollan et al.'s aim in developing the theory of distributed cognition was to ``provide new insights for the design of conceptually meaningful tools and work environments'' \cite{hollan2000}. Our findings showed that even though collaborative friction arose from domain gaps, UXPs saw more common ground when communicating with technical stakeholders after tinkering with simple AI models. Given this, we ask: \textit{what do conceptually meaningful tools and environments for collaborative design of human-centered AI look like?} These tools and environments should take into account the distributed and diverse modes of expertise in the process (e.g., AI expertise, UX expertise) while respecting the individual communities of practice, as indicated by findings from Section \ref{s:findings-tinkering}. 
The opportunity, then, lies in facilitating the fluid movement of cognitive processes and artifacts across bodies of expertise. 

We see potential in AutoML systems to act as a boundary object during collaboration. We use AutoML systems here to refer to platforms such as Teachable Machine that have both support for AI experts (e.g., code export, standard evaluation metrics) and non-experts (e.g., no-code GUI interface, high level of automation). In the current AutoML landscape, we still have little consensus on who AutoML benefits---providers of AutoML see it as a way to lower the barrier for AI non-experts to create robust models \cite{google-aml, ibm-aml, amazon-aml}, whereas prior studies have revealed that AI experts use them to prototype models and test technical hypotheses \cite{crisan2021automl, xin2021whither}. This lack of clarity has led to a default assumption that the path to advancing AutoML is one of increasing automation \cite{karmaker2021automl}. However, we offer another perspective: AutoML can benefit multiple user groups simultaneously is a reason for the lack of a clearly defined user group for AutoML. We specifically envision \textit{collaborative AutoML platforms} to be spaces where ideas and artifacts are shared across boundaries in the backdrop of technical infrastructure optimized for rapid prototyping. For example, UXPs may suggest data labels for supervised datasets based on user research, which AI practitioners can then use to source data. Similar to model tinkering in Teachable Machine, UXPs may draft up proof-of-concept models using AutoML, on which the AI team and product managers can leave comments about feasibility or resource restraints. AI practitioners may then choose to wrangle data and develop models outside AutoML using their preferred tools and environments, but when evaluating the trained models, diverse stakeholders can be brought back together to contribute a critical ``sanity check'' on outputs before the model is deployed. 


We note here the importance of involving one group of experts in particular with this collaborative AutoML approach: domain experts. These are individuals with deep knowledge of a domain in which the technology will be used, but may not necessarily be well-versed in UX or AI. In the case of our study, domain experts include nutritionists and dietitians. Several participants mentioned during the study warned users they should always consult a nutritionist or medical professional before accepting advice from the AI's output, and that working with a domain expert during the design process can yield nuanced design considerations they may otherwise miss on their own. However, surveys of industry UX practice show that collaborations with domain experts is still relatively uncommon \cite{feng2023collaboration}. In situations where lowering the barrier for collaboration does not increase collaboration rates due to deeply-established organizational and cultural norms, developing a richer understanding of how UXPs can effectively communicate AI as a design material to stakeholders becomes all the more important in ensuring a high quality of collaboration whenever it is available.


\subsection{Embracing Stochasticity in Design}
A key problem that emerges as UX and technical stakeholders collaborate more closely is the fundamental difference in the representation of success. In AI, models are evaluated on standard metrics such as classification accuracy, F1 score, and ROC AUC. In UX, designs are evaluated on user feedback: do they satisfy pain points and meet user needs? We see in Section \ref{s:accuracy} that UXPs apply the latter mindset to evaluating AI models. These two divergent evaluation strategies may prevent a \textit{lingua franca} to be established during collaboration. Current design tools have yet to mitigate this challenge due to their inability to support stochastic prototyping. Modern AI leverages statistical methods and therefore exhibits probabilistic, non-deterministic behaviour \cite{russell2010artificial}, but current design tools only enable the creation of deterministic user flows \cite{invision-user-flows}. This presents a problem when conducting user research with AI-enabled interfaces. If a defining property of the AI cannot be accurately communicated to users, they are unable to provide meaningful feedback on the design material. 

\begin{figure*}[h]
    \centering
    \includegraphics[width=1\textwidth]{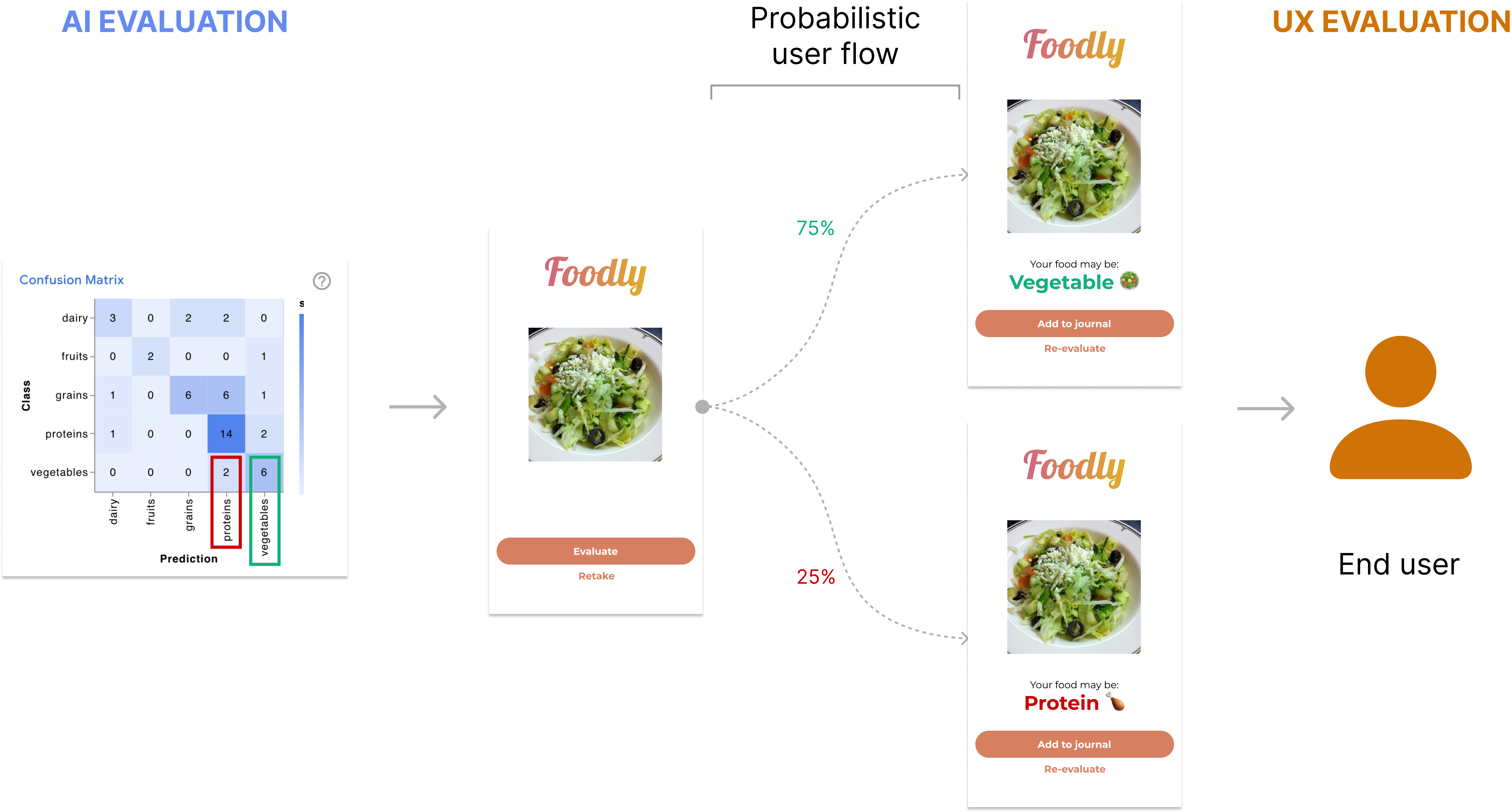}
    \caption{Probabilistic user flows are informed by AI-centric model evaluations while enabling user-centric evaluations, acting as a bridge between the two.}
    \label{f:puf}
    \Description{A confusion matrix on the left indicates a 75\% accuracy rate for a model for the vegetable label: 6 of 8 samples were correctly classified as vegetable, while the remaining two were classified as protein. A probabilistic user flow to the right shows an app screen with an image of a plate of vegetables branching off into two screens as the next stage in a user flow. The top of the two has 75\% chance of appearance and displays the correct classification of vegetable. The bottom one has 25\% chance of appearance and displays the incorrect classification of protein. A user icon is situated towards the right of these two screens, indicating exposure of these probabilistic screens to an end user.}
\end{figure*}

We propose \textit{probabilistic user flows} to not only simulate the stochastic nature of AI-enabled UIs, but also bridge AI-centric and user-centric evaluations. Instead of defining deterministic flow paths from one screen to another, probabilistic user flows enable UXPs to specify set flow paths with probabilities of entering each path based on evaluations from technical AI workflows. We can use a scenario from our study as a motivating example. Suppose a UXP would like to collect user feedback on their designs based on the 5-class food classification they trained in Teachable Machine, and the model's performance happens to be identical to the one shown in Fig. \ref{f:evaluations}. The accuracy evaluation of the model can be seen in the table and confusion matrix under the ``Accuracy in Advanced menu'' label in Fig. \ref{f:evaluations}. A UXP can craft a probabilistic user flow to simulate the behaviour of the model with a specified accuracy: when the user inputs an image with a ground truth label of \textsc{Vegetables}, there is a 75\% chance they will be taken to a screen with the model correctly identifying that the image is indeed one of vegetables and a 25\% chance they will be taken to a screen with an incorrect evaluation. This will allow the user to experience the model's probabilistic behaviour directly in the prototype and provide the feedback necessary for design evaluation. Moreover, the UXP can now leverage the standard AI evaluation of the model (class-based accuracies) directly in the UX evaluation of a design enabled by that model. We illustrate this example with a model's confusion matrix in Fig. \ref{f:puf}.

Probabilistic user flows, however, may quickly result in complex, non-linear user flows that overwhelm designers' workspaces with mathematical clutter. Efficient information management then becomes vital. If such user flows were to be implemented in design tools, it is important to explore information organization techniques along flow paths, which can include ``screen families'' to associate and collapse sets of screens bounded by the same probability vector, or moving the editing of probabilistic user flows to a separate environment (e.g., a modal popup).

%% file: sections/limitations-fw.tex
Our study was kept to around 2 hours to minimize participant fatigue while fitting in all of our planned components (tutorial, main design activity, interview). Due to this time constraint, the main design activity in our study only lasted for around 75 minutes. This may have caused participants to sacrifice quality for speed in their work. Although a few participants finished early, most used the entire 75 minutes and expressed a desire to polish up their work if they were given more time. Additionally, this study only included the perspective of UXPs. In practice, UXPs may frequently collaborate with non-UX team members and capturing only UX perspectives in those cases yields a narrative that is not representative of the broader team mentality. To address both of these limitations, future work may find it fruitful to conduct a longitudinal ethnography of interdisciplinary teams working within industry organizations to design and build AI interfaces. The additional perspectives in such as a study can initiate rich conversations with findings from our study, allowing new implications on collaboration to be surfaced.

This study focused on only the second half of the diamond design process---participants were given a well-defined design problem (along with accompanying resources such as user research insights and personas) and were asked to ideate and present a solution. That is, we assume AI is an appropriate design material to work with for the given problem. However, it is still an open question of how UXPs can arrive at such a conclusion, or a contrary one, when defining the design problem itself. Future work may choose to focus more on the first diamond in the design process and investigate how UXPs reason about whether AI is a suitable material based on their user research and synthesis.

Finally, this study was centered around a narrow application of AI. Food is a universally understood concept, and although the activity could benefit from UXPs' collaboration with domain experts (e.g., nutritionists), participants were able to quickly grasp and reason about the design problem. In some application areas of AI, such as medical diagnosis, domain-specific knowledge is indispensable. Inclusion of domain experts in future studies alongside UXPs can present unique and important case studies for design communication. The rise of natural language AI interfaces due to the success of large language models may also shift the nature of tinkering with the design material. While our study only included traditional classification models, future research can extend a similar approach to interactive playgrounds for generative AI.

%% file: sections/conclusion.tex
As AI becomes an increasingly approachable design material for UXPs, we asked: \textit{how do UXPs communicate AI to other stakeholders as a design material within their practice?} We conducted a contextual inquiry in the form of a design activity with 27 UXPs to answer this question. We found that although it is currently uncommon in industry settings, increasing the agency of material fabrication for UXPs while designing an AI-enabled interface increased their high-level understanding of basic AI concepts, the benefits and risks of applying ML to their designs, and how they can effectively collaborate with non-UX collaborators going forward. Based on our findings, we propose 3 implications for scaffold interdisciplinary work at the crossroads of UX and AI: applying the concept of \textit{fidelity} from prototyping to AI models as a sensitizing concept, extending AutoML platforms along a collaborative dimension to act as boundary objects, and incorporating \textit{probabilistic user flows} in design tools. We hope our work can illuminate the path for more collaborative conversations about the use of AI as a design material and, more broadly, a human-centered technology.